\documentclass{emulateapj}

\usepackage{apjfonts}

\journalinfo{Astrophysical Journal Letters}
\slugcomment{Received 16 Jun 2006; Accepted 24 Jul 2006; In press}

\begin{document}

\title{Dynamo Action in the Solar Convection Zone and Tachocline: Pumping and
  Organization of Toroidal Fields}

\author{Matthew K. Browning$^1$, Mark S. Miesch$^2$, Allan Sacha Brun$^{3}$, 
and Juri Toomre$^4$}

\affil{$^1$ Astronomy Department, University of California at Berkeley, 601
  Campbell Hall, Berkeley CA 94720-3411, matthew@astro.berkeley.edu,\\
$^2$High Altitude Observatory, NCAR, Boulder, CO 80307-3000, \\
$^3$DSM/DAPNIA/SAp, CEA Saclay, 91191 Gif sur Yvette, France, \\
$^4$JILA and Department of Astrophysical and Planetary Sciences,
University of Colorado, Boulder, CO 80309-0440}

\begin{abstract}
We present the first results from three-dimensional spherical shell simulations of
magnetic dynamo action realized by turbulent convection penetrating
downward into a tachocline of rotational shear.  This permits us to assess
several dynamical elements believed to be crucial to the operation of the
solar global dynamo, variously involving differential rotation resulting
from convection, magnetic pumping, and amplification of fields by
stretching within the tachocline.  The simulations reveal that strong axisymmetric
toroidal magnetic fields (about 3000 G in strength) are realized within the
lower stable layer, unlike in the convection zone where fluctuating fields
are predominant.  The toroidal fields in the stable layer possess a
striking persistent antisymmetric parity, with fields in the northern
hemisphere largely of opposite polarity to those in the southern
hemisphere.  The associated mean poloidal magnetic fields there have a
clear dipolar geometry, but we have not yet observed any distinctive
reversals or latitudinal propagation.  The presence of these deep magnetic
fields appears to stabilize the sense of mean fields produced by vigorous
dynamo action in the bulk of the convection zone.

\end{abstract}


\section{Elements of Solar Dynamo}

The solar global dynamo responsible for the observed 22-year cycles 
of magnetic activity is likely to require several dynamical processes
operating at differing sites within the convection zone and below its
base.  The primary elements  were first discussed within the conceptual 
``interface dynamo'' proposed by Parker (1993) and then explored by Charbonneau 
\& MacGregor (1997).  The key dynamical processes involve (a) the generation 
of magnetic fields by the intense turbulence influenced by rotation within the 
deep solar convection zone, (b)~the transport or pumping of these fields downward 
into the tachocline of shear at the base of this zone, (c) the stretching of
fields there by differential rotation to form strong toroidal fields, and
(d) the magnetic buoyancy instability of such structures leading to field
loops ascending toward the surface.  Fully self-consistent magnetohydrodynamic 
(MHD) simulations of the complete solar global dynamo are not yet feasible, given 
the vast range of dynamical scales involved in these turbulent processes.  However, 
many of the elements have now begun to be studied singly with reasonable fidelity 
using three-dimensional numerical modelling, often turning to localized planar 
domains to obtain sufficient spatial resolution within the simulations (see reviews 
by Ossendrijver 2003, Fan 2004, Charbonneau 2005, Miesch 2005).  

Yet some aspects require using full spherical shell geometries in the modelling. 
Early simulations of dynamo processes in spherical shells provided insights into
the coupling of convection, rotation and magnetism, but were limited by spatial
resolution (e.g., Gilman 1983; Glatzmaier 1985a, b). 
Recent studies of the interaction of
turbulent convection with rotation (e.g., Miesch et al. 2000; Brun \&
Toomre 2002; Miesch, Brun \& Toomre 2006) now yield global differential rotation
profiles in close accord with helioseismic deductions (e.g., Thompson et
al. 2003) in the bulk of the convection zone.  Similarly, related MHD
modelling of dynamo processes in such deep shells of turbulent convection
has revealed that strong magnetic fields are produced without diminishing
the differential rotation significantly (Brun, Miesch \& Toomre 2004,
hereafter BMT04).  Fluctuating magnetic fields with strengths of order
5000 G are realized that possess complex structures, with radial fields
concentrated in downflow lanes and toroidal fields appearing as twisted
ribbons extended in longitude.  However, the associated mean fields are
relatively weak and do not exhibit the systematic latitudinal propagation
or periodic polarity reversals seen in the Sun.  Whereas BMT04 only considered 
dynamical element (a), we report here on also incorporating elements (b) and (c) 
into global simulations of dynamo action achieved by turbulent convection 
able to penetrate downward into a tachocline of rotational shear. Such
modeling is the next step in refining our intuition about the operation of
the interface dynamo.

\section{Convective Shell with Penetration and Forced Tachocline}

We conduct three-dimensional nonlinear MHD simulations of convection and dynamo
action in a rotating spherical shell using the anelastic spherical harmonic 
pseudospectral code ASH (see Clune et al. 1999 and BMT04).  ASH is based 
upon a large-eddy simulation (LES) approach involving subgrid-scale
(SGS) modelling of unresolved turbulent processes.  Our studies here extend the 
modelling of BMT04 by allowing penetration of the convection downward into a region of 
stable stratification, where the pronounced differential rotation maintained 
by the convection is forced to vanish, thereby establishing a tachocline of 
shear.  Our computational domain extends over 0.62--0.96 $R$, where $R$ 
is the solar radius, spanning the bulk of the solar convection zone and part of 
the radiative interior below.  Solar values are used for the rotation rate 
and luminosity, and the initial stratification is obtained from a one-dimensional solar 
structure model.  We adopt a softer subadiabatic stratification in the lower 
stable zone to ease our numerical resolution of internal gravity waves produced 
by the overshooting motions (see Miesch et al. 2000).  

The tachocline as revealed by helioseismology is a narrow transition
boundary layer between the differentially rotating convection zone (fast
equator, slow poles) and the uniformly rotating radiative interior.  Its
discovery motivated the interface dynamo paradigm.  The detailed structure
of the tachocline may be determined variously by anisotropic turbulent
mixing processes, magnetic stresses, and production of internal gravity
waves; the equilibration times likewise continue to engender debate (see
review by Miesch 2005).  Despite uncertainties, a tachocline of rotational
shear is crucial to element (c), and thus we seek to impose one here in two
complementary ways.  First, we introduce a drag force upon the axisymmetric
velocities (relative to our rotational frame) to force them to vanish at the
base of the computational domain.  We accomplish this smoothly with a
hyperbolic tangent (of width 0.01 $R$ centered at 0.66 $R$ in the stable
layer) so that motions within the bulk of the convective envelope are
unimpeded.  Without this forcing, the differential rotation established
self-consistently within the convection zone would imprint itself
upon the radiative zone through viscous and thermal diffusion.  Second, we
impose within the overshooting region a weak latitudinal entropy variation
to emulate the coupling between the convective envelope and the
radiative interior through thermal wind balance within the solar
tachocline.  This is achieved by thermal forcing, involving a monotonic
increase in temperature of about 6 K from equator to pole at the base of
convection zone (over a width of 0.02 $R$), much as discussed by
Miesch, Brun \& Toomre (2006, hereafter MBT06).  The thermal forcing aids
in achieving a strong latitudinal differential rotation within the
convection zone, which otherwise would be diminished by coupling to the
radiative interior.  This coupling is much stronger in our simulations than
in the Sun because the viscous and thermal diffusivities are far larger and
the overshoot region is wider.

The domain boundaries are assumed to be impenetrable and free of viscous
stresses.  The lower boundary is assumed to be a perfect conductor and the
radial entropy gradient is fixed there.  At the upper boundary the magnetic
field within the domain is matched to an external potential field and the
entropy is fixed.  The simulations were initiated by adding a weak seed
field to a progenitor penetrative hydrodynamic simulation.  The seed field
was toroidal and confined to the bulk of the convection zone.  For the
present model, we employ a horizontal resolution of $N_\theta = 512$,
$N_\phi = 1024$, and a stacked Chebyshev expansion in the radial dimension
with $N_r = 98$.  The effective SGS viscosity $\nu$ is equal to $6 \times
10^{12}$ cm$^{2}$ s$^{-1}$ at the top of the domain and decreases with
depth as $\overline{\rho}^{-1/2}$, where $\overline{\rho}$ is the mean
density varying by a factor of 46 across the domain.  The thermal
($\kappa$) and magnetic ($\eta$) SGS diffusivities have a similar profile,
with fixed Prandtl and magnetic Prandtl numbers of Pr = $\nu/\kappa = 0.25$
and Pm = $\nu/\eta = $ 8 respectively. The large Pm here reflects
unresolved turbulent mixing processes, and such a value allows efficient
dynamo action with tractable numerical resolution.  In the Sun, Pm based on
microscopic processes is much smaller, but the relationship between it and
the effective turbulent diffusivities is uncertain.  At small Pm, the
critical magnetic Reynolds number needed for dynamo action increases
considerably (Boldyrev \& Cattaneo 2004; Schekochihin et al. 2005),
rendering simulations much more computationally demanding.  The field
strengths reported here are likely sensitive to the Pm chosen.

\section{Dynamo Action Yielding Organized Magnetic Fields}

The convective flows in the simulation are intricate and highly time
dependent.  Fairly broad, low-amplitude upflows coexist with fast,
concentrated downflows, which are characteristic of turbulent
compressible convection (e.g., Brun \& Toomre 2002; BMT04; Brummell et
al. 2002).  Some downflow lanes persist as coherent structures
for extended intervals.  The flows are sampled in Figure 1$a$, which
shows a global mapping of the radial velocity $v_r$ on a spherical surface
within the convection zone (at $r=0.84R$).  Most of the downflow plumes
visible there extend downward and overshoot into the underlying radiative
region, where they are buoyantly braked.

\begin{figure}[hpt]
\center
%\epsscale{1.25}
%\plotone{f9.eps}
\includegraphics[width=3.4in, trim= 0 0 320 600]{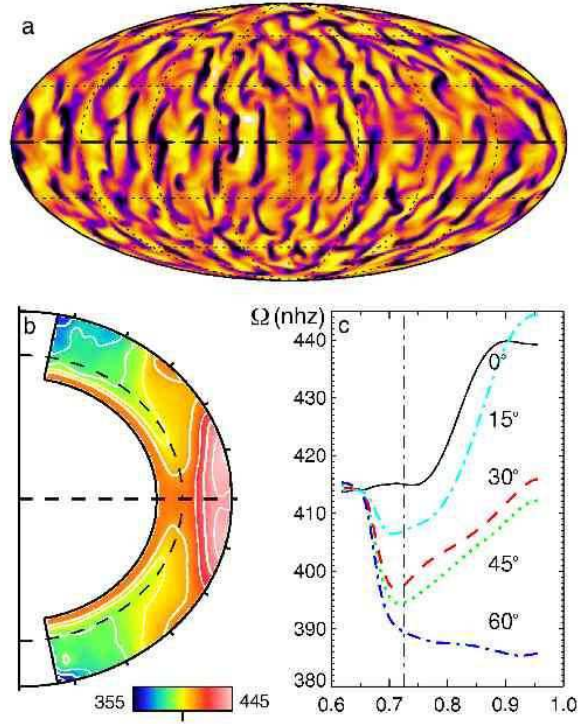} % {p6_vr_diffrot3f_sm.ps}
\caption{\label{vr_diffrotn} ($a$) Global view (in Mollweide projection) at
one instant of radial velocity $v_r$ on a spherical surface at mid-depth
($r=0.84R$) in the convection zone, with upflows bright and downflows
dark.  Equator is shown dashed.  Angular velocity $\Omega$ averaged in time and in longitude,
showing ($b$) contours in radius and latitude
and ($c$) variation with proportional radius along specified latitudinal cuts.}
\end{figure}

\begin{figure*}[htp]
\center
%\epsscale{1.25}
%\plotone{f9.eps}
\includegraphics[width=7.0in, trim= 0 32 32 580]{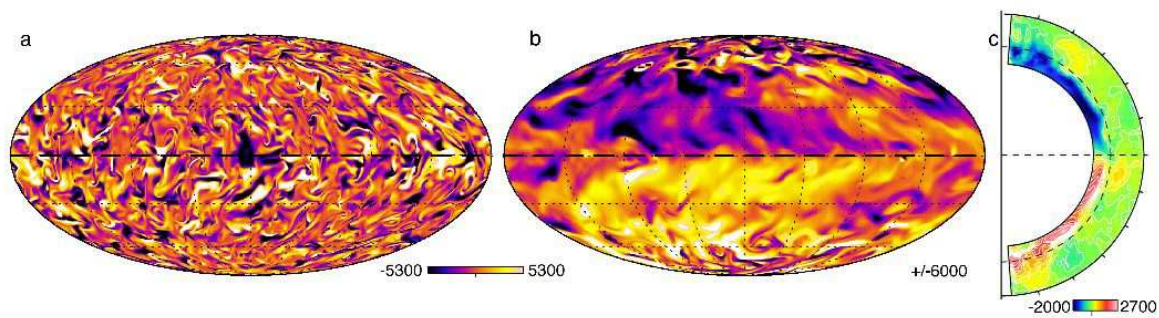} %{bphi_mollw_3panel3_sm.ps}
\caption{\label{bphi_3panel} Toroidal magnetic fields realized in the
  convective envelope and underlying radiative region.  Mollweide
projections at one instant of longitudinal field $B_{\phi}$ on a spherical surface
  ($a$) at mid-depth in the convection zone ($r=0.84R$) and ($b$) in
the stable zone ($r=0.67R$).  ($c$) Contours in radius and latitude of
$B_{\phi}$ averaged in time (over an interval of 220 days) and in
longitude.  Field strengths are in G.}
\end{figure*}

Strong differential rotation is established by the convection.  Figure 1$b$
shows the mean angular velocity $\Omega$ (relative to the rotating frame)
as a contour plot in radius and latitude, and Figure 1$c$ its variation
with radius along selected latitudinal cuts.  Within the bulk of the
convection zone, there is a prominent decrease in $\Omega$ from equator to
pole.  Near the upper boundary, the angular velocity
contrast $\Delta \Omega$ (between equator and 60\degr ) is 55 nHz or 13\% of the frame rotation rate.  In
comparison to MBT06 that had no penetrative region, the $\Delta \Omega$
here is reduced and $\Omega$ is more constant on cylinders aligned with the
rotation axis.  Below the base of the convection zone (at $0.73R$) where
the stratification is subadiabatic, $\Omega$ adjusts rapidly to uniform
rotation within a pronounced tachocline of shear.  The overall fast equator
and slow pole behavior is in keeping with the $\Omega$ profiles deduced
from helioseismology (e.g., Thompson et al. 2003), but this model possesses
more radial shear within the bulk of the convection zone than in the Sun.
We have clearly realized a tachocline of strong radial and latitudinal
shear, which plays a major role in the organization of large-scale magnetic
fields.

Prominent magnetic dynamo action is achieved within this simulation.  Much
as in BMT04, the seed fields are rapidly amplified, with strong fluctuating
magnetic fields ($\sim$ 3000 G rms) and weaker mean toroidal and poloidal
fields (of time-averaged strengths $\leq$ 300 G) achieved within the bulk
of the convection zone.  Over the 2800 simulated days (about 100 rotation
periods) studied after the field strengths have largely equilibrated, the
dynamo action is persistent, with overall magnetic energy sustained at
about 40\% of the convective kinetic energy.  Figure $2a$ shows a snapshot
of the longitudinal field $B_{\phi}$ at mid-depth ($r=0.84R$), revealing
complex structure on many scales, with no evident polarity preferences.
The radial and latitudinal fields $B_r$ and $B_{\theta}$ also possess
intricate and highly variable structures, tracing the convective flow
realized in the bulk of the convection zone. In the underlying stable
region (at $r=0.67R$), Figure 2$b$ shows that the magnetic field has been
decidedly organized by the rotational shear, with the longitudinal field
there stretched into large toroidal structures that extend around much of
the domain.  Equally striking is that $B_{\phi}$ fields show antisymmetric
parity in the stable region, with opposite signs to those in the northern
and southern hemispheres. The organized nature of strong toroidal fields in
the stable region is confirmed in Figure 2$c$ showing the time-averaged
axisymmetric $B_{\phi}$. Here the opposite polarities of the mean
toroidal fields are evident in the two hemispheres, as contrasted to the
weak and patchy mixed-polarity structures within the convection zone.  The
time-averaged axisymmetric fields in the stable region attain strengths of
order 3000 G, or about ten times stronger than the mean $B_{\phi}$ in the
convection zone.  Whereas fluctuating (non-axisymmetric) fields dominate in
the convection zone, the magnetic energy in the mean toroidal field is
about three times larger than the fluctuating magnetic energy within the
tachocline.  The strong toroidal field established in the stable region is
accompanied by a largely dipolar poloidal field.

Neither the organization of the magnetic field below the convection zone
into predominantly axisymmetric toroidal fields, nor the strong parity
selection exhibited here, appear to be transient effects. These attributes
arose rather quickly ($\sim 200$ days) after the introduction of the
forced tachocline, and have persisted for as long as we have continued the
calculations.  A sense of the evolution and spatial distribution of the
mean fields in our simulation is provided by Figure 3 showing the radial
and temporal variations of the axisymmetric toroidal field $B_{\phi}$
sampled at latitude $-30$\degr.  The strongest toroidal fields there are
clearly attained in the stable region ($r/R <$ 0.73), with some modulations
in amplitude but no changes in polarity.  Within the convection zone, both
the strength and polarity of the weaker toroidal field are more variable.
Further investigation reveals that the axisymmetric dipole field both in the
convection zone and in the stable region has not changed its overall
polarity during the course of our simulation. This is in sharp contrast to
the evolution of the mean dipole component in BMT04, which considered the
convection zone in isolation, where the dipole flipped at irregular
intervals of less than 600 days.  This suggests that the presence of a
reservoir of strong toroidal fields with persistent polarity in the stable
region is serving to stabilize the mean poloidal field realized in the
convection zone.

\section{Building Strong Toroidal Fields}

The predominantly axisymmetric nature of the magnetic fields in the stable
region may be understood in two complementary ways.  First, the lack of
significant non-axisymmetric motions there precludes the generation of
strong fluctuating fields like those realized in the convection zone. Thus
non-axisymmetric (fluctuating) fields must either diffuse in from the convection zone or
be transported downward by overshooting motions.  Second, Spruit (1999) has
argued that any such fluctuating fields will, in the presence of
well-defined angular velocity gradients, be quickly expelled from the
system through reconnection between neighboring magnetic surfaces.  He
estimates that this process acts on a time scale $\tau_{\Omega} \sim (3 r^2
\pi^2/\eta \Omega^2 q^2)^{1/3}$, with $q$ a measure of the rotational
shear. In the Sun, this estimate yields time scales of order 100 years, but
in our far more diffusive simulation, $\tau_{\Omega}$ is about a year.
Thus fluctuating fields pervading the radiative region are fairly
quickly erased; they are not replenished from above on a comparable time
scale.

\begin{figure}[hpt]
\center
%\epsscale{1.25}
%\plotone{f9.eps}
\includegraphics[width=3.5in, trim= 160 380 72 120]{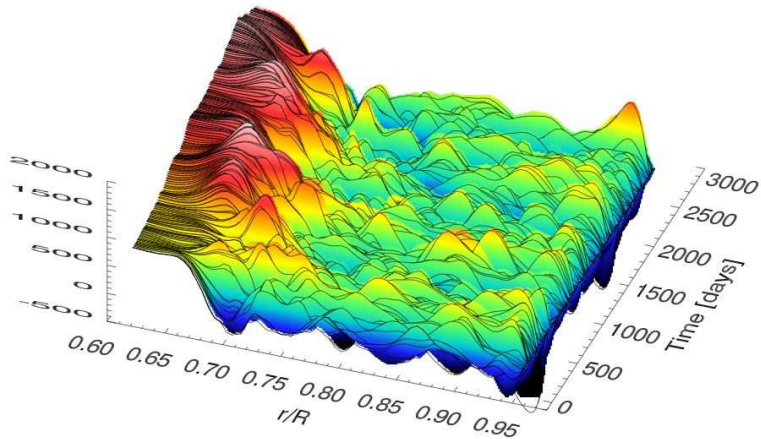} %{Bphi_radiustime_sm.ps}
\caption{\label{bphi_radtime} Temporal evolution and spatial distribution
of axisymmetric toroidal magnetic fields.  Variation with radius and time
of $B_{\phi}$ at a latitude of $-30$\degr \ , with tall peaks (bright tones)
corresponding to large positive amplitudes.  The strongest fields are
realized in the stable region ($r/R < 0.73$), where the polarity of the fields is also
remarkably stable over the 2500 days sampled here.}
\end{figure}

The strength of the mean toroidal field below the convection zone is likely
strongly influenced by the enhanced diffusive terms and spatially extended
tachocline in our simulation.  In the kinematic regime, the strength of
toroidal fields $B_t$ generated in time $t$ by stretching of poloidal
fields $B_p$ due to radial shear is of order $B_t/B_p \sim \Delta v_{\phi}
t / \Delta r$, with $\Delta v_{\phi}$ the radial differential rotation
velocity across a length $\Delta r$ below the convection zone.  Thus the
strength of the toroidal field depends both on the angular velocity
contrast and on how abrupt that contrast is; a narrower tachocline might
then yield stronger fields.  This estimate suggests that in our simulation,
a 10 G poloidal field could be stretched to yield 1 kG toroidal fields in
about a year.  The magnetic diffusion time scale at the same depth is about
3 years, so fields can be amplified before they are diffused away.  The far
smaller magnetic diffusion and possibly narrower tachocline in the real
solar interior may well lead to even greater amplification of the toroidal
field there.

The realization of antisymmetric toroidal field parity within the stable
zone is a striking property. It appears to be a robust feature but its origins
are currently unclear. The generation of such parity would be expected if a
dipolar poloidal field were to be sheared by differential rotation, but no
such dipole seeds were introduced. Rather, the initial seed field was toroidal
and included both symmetric and antisymmetric components. The emergence
of a dipolar mean field may be a consequence of large-scale
self-organization processes such as those associated with the inverse cascade
of magnetic helicity in MHD turbulence or the $\alpha$-effect of mean-field
dynamo theory. Such processes may occur within the convection zone
where Coriolis forces induce opposite kinetic helicity in the northern and
southern hemispheres (as in BMT04). Poloidal fields thus generated may then
be pumped into the tachocline where they are stretched into toroidal structures
by the shear and amplified. The reservoir of toroidal field thus established
within the tachocline may then feed back on field generation in the convection
zone, providing the seed for new poloidal field. We plan to assess each of these
processes in future work.

In summary, these are the first global three-dimensional MHD simulations of
turbulent solar convection to allow penetration into a forced tachocline of
rotational shear.  We have thus made contact with several of the dynamical
processes thought to be essential in the operation of the global solar
dynamo.  The combined action of convection and an organized tachocline has
here yielded magnetic fields that possess several distinctive properties.
The significant toroidal fields realized beneath the convection zone, the
antisymmetric parity displayed by those fields, and the persistence of a
single polarity for multiple years, are reminiscent of the highly organized
magnetism that appears as sunspots at the solar surface.  Further work to
test the robustness of these features has been initiated and will be reported.

This work was partly supported by NASA through Heliophysics Theory Program
grant NNG05G124G, and by NSF through an Astronomy and Astrophysics
Postdoctoral Fellowship (AST-0502413).  The simulations were carried out
with NSF PACI support of NCSA, SDSC and PSC, NASA support of Project
Columbia, as well as the CEA resource of CCRT and CNRS-IDRIS in France.

\end{document}